\documentclass[aps,prb,reprint,superscriptaddress]{revtex4-2}
\usepackage{bm}
\usepackage{amsmath, amssymb, amsfonts, latexsym}
\usepackage{lipsum}
\usepackage{graphicx}
\usepackage{xcolor}

\newcommand{\bvec}[1]{\mbox{\boldmath $#1$}}

\begin{document}

\title{Field-induced antiferromagnetic transition in CeIrIn$_5$}

\author{Y.~Tokunaga}
\email[]{tokunaga.yo@jaea.go.jp}
\affiliation{ASRC, Japan Atomic Energy Agency, Tokai, Ibaraki 319-1195, Japan}

\author{M.-T.~Suzuki}
\affiliation{Department of Materials Science, Graduate School of Engineering, Osaka Metropolitan University, 1-1 Gakuen-cho, Naka-ku, Sakai, Osaka 599-8531, Japan}
\affiliation{Center for Spintronics Research Network, Graduate School of Engineering Science,
Osaka University, Toyonaka, Osaka 560-8531, Japan}

\author{S.~Kr\"{a}mer}
\affiliation{
Laboratoire National des Champs Magn\'{e}tiques Intenses, LNCMI-CNRS (UPR3228), EMFL, \\
Univ.\,Grenoble Alpes, Univ.\,Toulouse, INSA-T, 38042 Grenoble Cedex 9, France}

\author{H.~Sakai}
\affiliation{ASRC, Japan Atomic Energy Agency Tokai, Ibaraki 319-1195, Japan}

\author{S.~Kambe}
\affiliation{ASRC, Japan Atomic Energy Agency Tokai, Ibaraki 319-1195, Japan}

\author{H.~Harima}
\affiliation{Graduate School of Science, Kobe University, Kobe 657-8501, Japan}

\author{D.~Aoki}
\affiliation{IMR, Tohoku University, Oarai, Ibaraki 311-1313, Japan}

\author{M.~Horvati{\'c}}
\affiliation{
Laboratoire National des Champs Magn\'{e}tiques Intenses, LNCMI-CNRS (UPR3228), EMFL, \\
Univ.\,Grenoble Alpes, Univ.\,Toulouse, INSA-T, 38042 Grenoble Cedex 9, France}

\author{I.~Sheikin}
\email[]{ilya.sheikin@lncmi.cnrs.fr}
\affiliation{
Laboratoire National des Champs Magn\'{e}tiques Intenses, LNCMI-CNRS (UPR3228), EMFL, \\
Univ.\,Grenoble Alpes, Univ.\,Toulouse, INSA-T, 38042 Grenoble Cedex 9, France}
\date{\today}

\begin{abstract}

We report low-temperature $^{115}$In nuclear magnetic resonance (NMR) study of the prototypical heavy-fermion compound CeIrIn$_5$ in high magnetic fields applied close to the crystallographic $c$ axis. For this orientation, a field-induced transition was previously reported to take place at about 28~T. Although we do not observe any change of the NMR spectrum above the transition, the intensity of the NMR lines drastically decreases as a consequence of a considerable shortening of the $T_2$ relaxation time. In addition, $1/T_1$ shows a pronounced maximum at the transition. Taking into account previous high-field de Haas-van Alphen results in conjunction with band-structure calculations, our NMR results are most naturally explained by the field-induced transition into an antiferromagnetic state with the propagation vector $\mathbf{Q} = (1/2, 1/2, 0)$ and magnetic moments aligned antiferromagnetically along the $c$ axis. This makes CeIrIn$_5$ a unique case where the application of the magnetic field induces an ordered state with moments antiferromagnetically aligned along the field direction.

\end{abstract}

\maketitle


The ground state of the intermetallic heavy-fermion compounds can be readily tuned by such control parameters as pressure, chemical composition, and magnetic field. The latter appears to be the most convenient as, contrary to pressure, it can be varied continuously, and, contrary to chemical doping, it does not induce disorder. In antiferromagnetic (AFM) heavy-fermion compounds, the magnetic field can be used to suppress magnetic order, driving them to a quantum critical point~\cite{Gegenwart2002,Harrison2007,Zeng2014,Jiao2015}. In addition, the magnetic field can sometimes induce reentrant superconductivity~\cite{Levy2005,Aoki2009,Knebel2019,Ran2019} or rather exotic phase transitions, such as Lifshitz transitions~\cite{Daou2006,Rourke2008,Pfau2013,Pourret2013,Bastien2016} or electronic nematicity~\cite{Ronning2017}.

CeIrIn$_5$ is one of the best-studied heavy-fermion materials. It is nonmagnetic and superconducting below the critical temperature $T_c = 0.4$~K~\cite{Petrovic2001}. The electronic specific heat coefficient is large, $\gamma = 750$~mJ/K$^2$mole~\cite{Movshovich2001}. Previous NMR measurements revealed a Knight shift anomaly at $T^*\sim 40$~K~\cite{Shockley2013,Shockley2015}. This temperature was suggested to correspond to the Kondo-lattice coherence temperature, below which the $4f$ electrons increasingly contribute to the itinerant heavy-electron state~\cite{Curro2004,Shirer2012}. This is indeed confirmed by low-temperature de Haas-van Alphen (dHvA) effect measurements~\cite{Haga2001,Capan2008,Aoki2016}, which revealed Fermi surfaces (FSs) whose topology is well accounted for by band-structure calculations treating the $f$ electrons as itinerant. The FSs consist mostly of quasi-two-dimensional $\alpha$ and $\beta$ sheets.

When a magnetic field is applied along or close to the $c$ axis, a phase transition occurs at $H^* \sim$~28~T. The transition manifests itself by clear anomalies in magnetic torque~\cite{Palm2003,Capan2008,Aoki2016}, resistivity~\cite{Capan2008}, specific heat~\cite{Kim2002}, thermoelectric power~\cite{Aoki2016} and ultrasound velocity~\cite{Kurihara2023} measurements. However, no anomaly was observed in either pulsed-field~\cite{Takeuchi2001} or steady-field~\cite{Aoki2016} magnetization measurements. High-field dHvA measurements~\cite{Aoki2016} revealed that most of the major FSs are preserved across the transition, while the largest $\beta_1$ frequency, originating from the $\beta$ sheet of the FS, completely disappears. Concurrently, a new low dHvA frequency with a strongly enhanced effective mass emerges. These observations were interpreted as evidence for a field-induced Lifshitz transition~\cite{Aoki2016}.

Field-induced Lifshitz transitions were previously observed in other heavy-fermion materials, such as CeRu$_2$Si$_2$~\cite{Daou2006} and YbRh$_2$Si$_2$~\cite{Rourke2008,Pfau2013,Pourret2013}. In these compounds, the Lifshitz transition is accompanied by a distinct anomaly in longitudinal magnetization, while in CeIrIn$_5$ there are no magnetization anomalies at $H^*$. On the other hand, the anomaly in magnetic torque, which is proportional to the transversal magnetization, suggests that the magnetic properties of CeIrIn$_5$ do change at $H^*$. This motivated us to further investigate this transition by means of NMR, which is a sensitive microscopic probe of magnetic phase transitions and spin dynamics.

In this Letter, we report high-field NMR measurements on a high-quality single crystal of CeIrIn$_5$ for a field applied at a few degrees from the $c$ axis. For this field orientation, the transition occurs at 30.5~T. Above the transition, the relaxation time $T_2$ shortens considerably implying a change of the spin dynamics at the transition. Furthermore, $1/T_1$ shows a pronounced maximum at the transition suggesting an enhancement of the spin fluctuations in the vicinity of the transition. The simplest explanation of these observations is a field-induced AFM transition. However, the NMR spectra do not change above the transition, which sets limitations on the possible magnetic structures. Furthermore, only one of them provides a consistent explanation of the previous high-field dHvA results.


\begin{figure}[htb]
\begin{center}
\includegraphics[width=8.2cm,keepaspectratio]{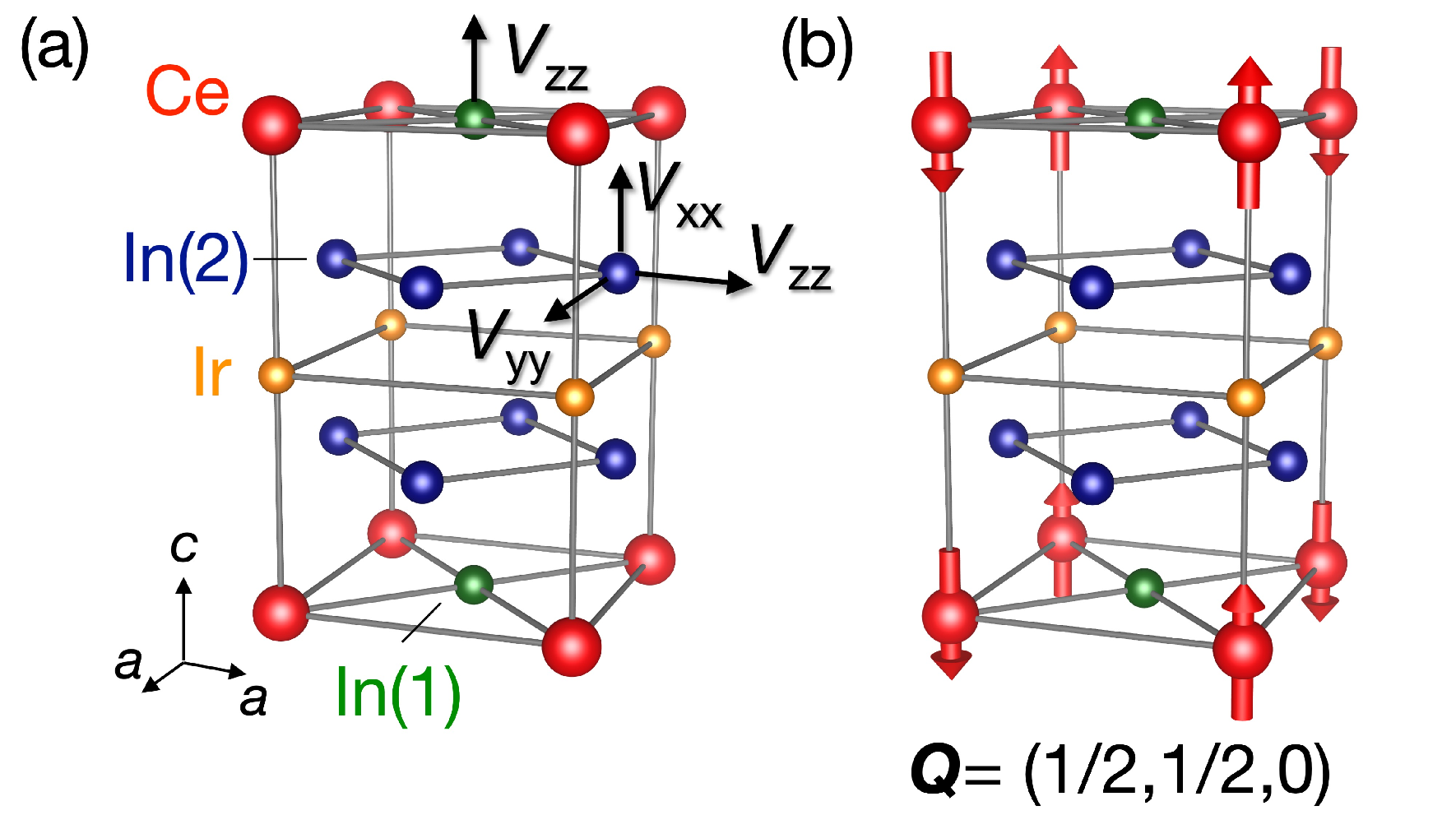}
\end{center}
\vspace{-5mm}
\caption{(a) Crystal structure of CeIrIn$_5$. Arrows indicate the principal axes of the EFG tensors for In(1) and In(2) sites. (b) The only high-field magnetic structures of CeIrIn$_5$ compatible with both the present NMR and previous dHvA~\cite{Aoki2016} results. See text for details.}
\label{f1}
\end{figure}

\begin{figure}[htb]
\begin{center}
\includegraphics[width=7.2cm,keepaspectratio]{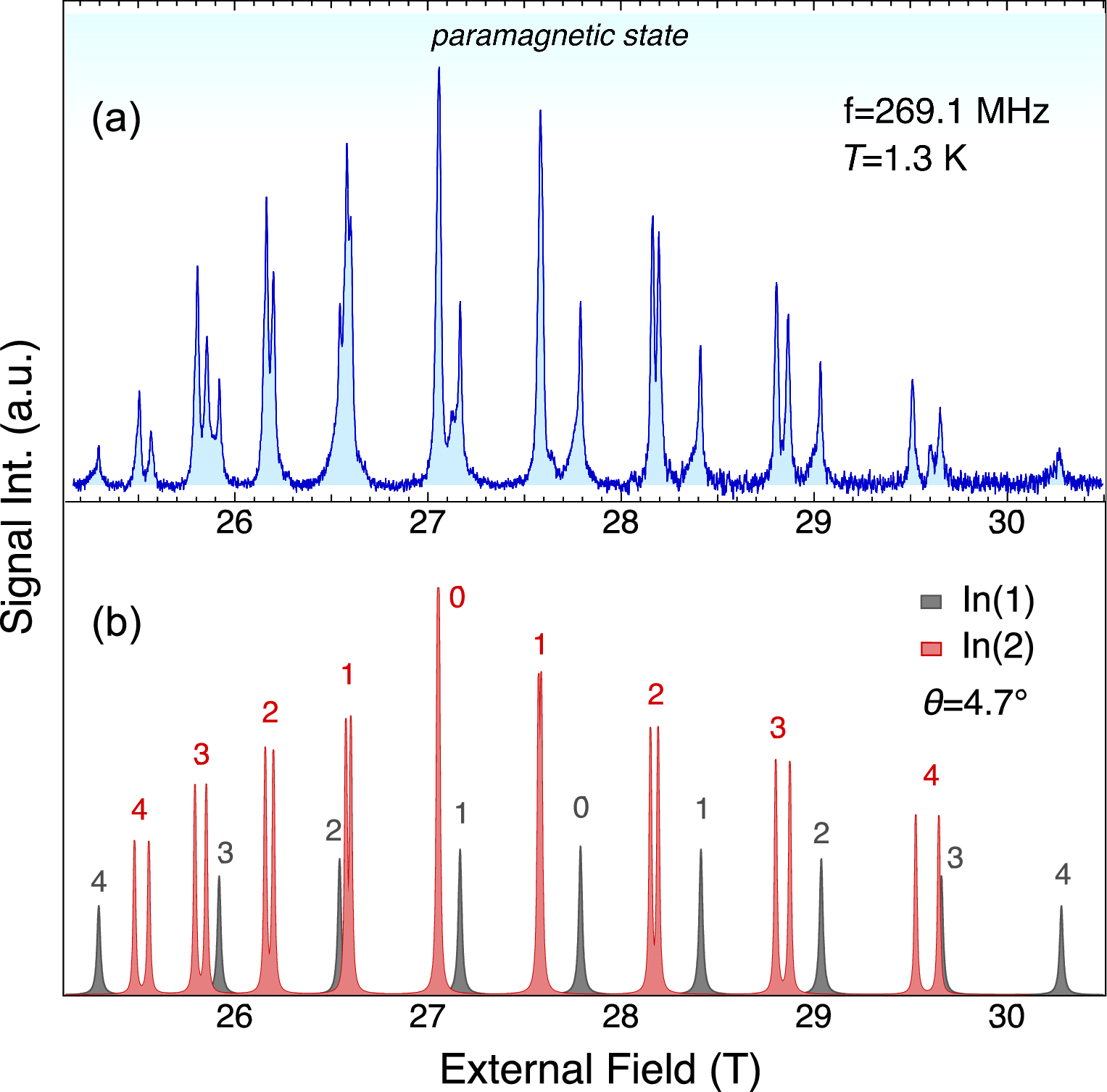}
\end{center}
\vspace{-2mm}
\caption{(a) Field-swept $^{115}$In NMR spectrum obtained at a fixed frequency of 269.1 MHz.  (b) Numerical simulation of the NMR spectra for In(1) and In(2) sites.  The simulation takes into account the effect of the field-dependent Knight shift shown in Fig.~\ref{f4}(a). As for the numbers labeled on each peak, zero denotes the center line, and 1–4 correspond to the first through the fourth satellite lines.}
\label{f2}
\vspace{-2mm}
\end{figure}

\begin{figure}[htb]
\begin{center}
\includegraphics[width=7.0cm,keepaspectratio]{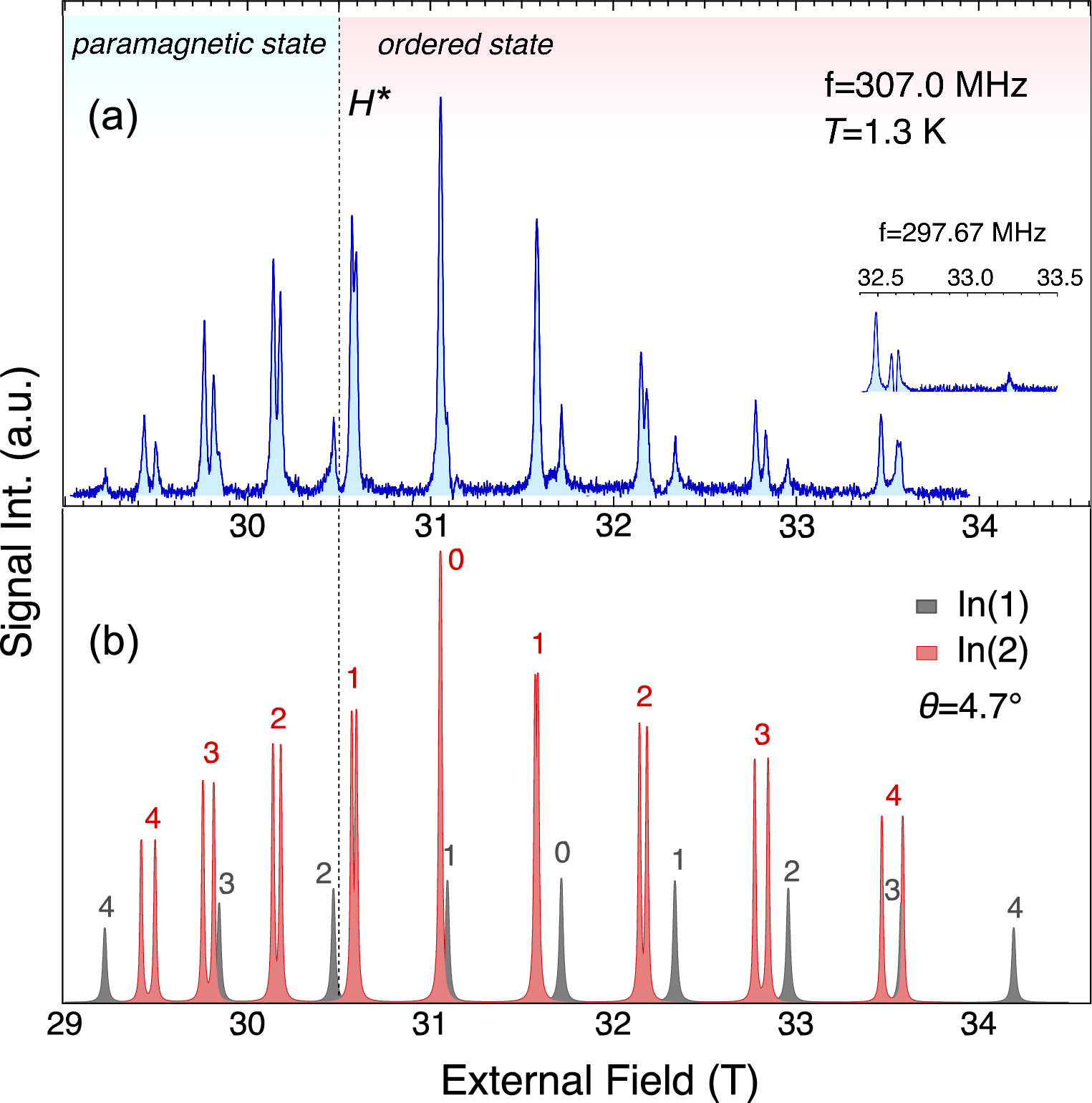}
\end{center}
\vspace{-2mm}
\caption{(a) Field-swept $^{115}$In NMR spectrum measured at a fixed frequency of 307.0 MHz. Since the highest available field was 34~T, the highest-field fourth satellite line of In(1) was recorded at a different frequency of 297.67 MHz. (b) Numerical simulation of the NMR spectra obtained using the same EFG parameters as those at lower fields.}
\label{f3}
\vspace{-2mm}
\end{figure}


All the high-field $^{115}$In-NMR measurements reported here were performed at a temperature of $T$ = 1.3~K in fields between 25 T and 34 T using a 24-MW resistive magnet at the LNCMI-Grenoble.  A high-quality single crystal of CeIrIn$_5$ with dimensions 4$\times$2$\times$0.7~mm was grown by the In self-flux method, the details of which are described in Ref.~\cite{Shishido2002}. CeIrIn$_5$ crystallizes in the tetragonal HoCoGa$_5$-type structure ($P4/mmm$) with two crystallographically inequivalent In sites: the locally tetragonal In(1) (1$c$ site) and the orthorhombic In(2) (4$i$ site) [Fig.~\ref{f1}(a)]. The principal axes of their electric field gradient (EFG) tensors $X$, $Y$, and $Z$ are illustrated in Fig.~\ref{f1}(a). Due to the local symmetry, the EFG tensor asymmetry parameter $\eta$, defined as $\eta=(V_{XX}-V_{YY})/V_{ZZ}$, is zero for In(1) and nonzero for In(2) sites.

Field-sweep NMR spectra at fixed frequencies were obtained by summing the Fourier transformation of the spin-echo signals recorded during stepwise field sweeps. The angle between the applied field and the crystallographic $c$ axis $\theta$ was determined by analyzing the NMR spectrum. For the spectrum analysis, we conducted numerical simulations based on the diagonalization of the total Hamiltonian matrix comprising the EFG and Zeeman terms. Under magnetic fields, each In site produces nine distinct NMR peaks due to quadrupolar splitting of the $^{115}$In nuclear spin $I$ = 9/2. Tilting of the magnetic field away from the $c$ axis further splits In(2) satellite peaks into two.

The field dependence of the NMR Knight shift, $K(H)$, was measured by tracking the peak frequency of the central transition for In(1) and In(2) sites as a function of field.
The NMR $1/T_1$ and $1/T_2$ were measured using a low-field (high-frequency) fourth satellite peak of In(2) sites.
To measure $1/T_1$, we employed the saturation-recovery method and evaluated it by fitting the recovery data $R(t)$ to a theoretical function of the $I=9/2$ nuclei. Similarly, $1/T_2$ was determined by monitoring the decay of the spin-echo intensity $I(\tau)$ as a function of the interval time $\tau$ between the $\pi/2$ and $\pi$ pulses, and then fitting $I(\tau)$ to an exponential function. Satisfactory fits with a single component of $T_1$($T_2$) for $R(t)$ ($I(\tau)$) have been achieved across the entire field range.


Figure~\ref{f2}(a) shows the field-swept NMR spectrum measured at a fixed frequency of 269.1 MHz. A number of sharp NMR peaks were observed in fields ranging from 25.2~T to 30.3~T. The numerical simulation shown in Fig.~\ref{f2}(b) confirmed that all the NMR peaks are well reproduced assuming $\theta=4.7^\circ$ and the EFG parameters $\nu_Q=6.083$ MHz and $\eta=0$ for In(1) and  $\nu_Q=18.167$ MHz and $\eta=0.465$ for In(2), respectively. Here, we successfully used the same EFG parameters as those obtained from a frequency-swept spectrum recorded at a lower field of 12.1~~T ($H \parallel c$)~\cite{Suppl}, providing \textit{a priori} a more reliable fit.

\nocite{Zheng2001,Kohori2001}

In Fig.~\ref{f3}(a), we present the field-swept NMR spectrum measured at 307.0 MHz, while its numerical simulation is shown in Fig.~\ref{f3}(b).  As discussed later,  both $T_1$ and $T_2$ measurements demonstrate a clear signature of a phase transition at $H^*$ = 30.5~T. Consequently, this NMR spectrum, spanning field values from 29 to 34~T, covers both paramagnetic ($H<H^*$) and ordered ($H>H^*$) phases. Nevertheless, we found that the positions of all the NMR peaks remain reproducible using the same EFG parameters as those at lower fields. Furthermore, there is no obvious broadening of the NMR peaks above $H^*$. These results suggest that the phase transition at $H^*$ is not accompanied by any structural change or anisotropic charge distribution (i.e. nematic order), and, moreover, does not induce a large internal field along the field direction. On the other hand, if we compare intensities of the corresponding In(2) satellite NMR lines above and below $H^*$, it is obvious that the intensities at higher fields are substantially reduced. As mentioned later, this reduction arises from a shortening of $T_2$ above $H^*$.

\begin{figure}[htb]
\begin{center}
\includegraphics[width=7.2cm,keepaspectratio]{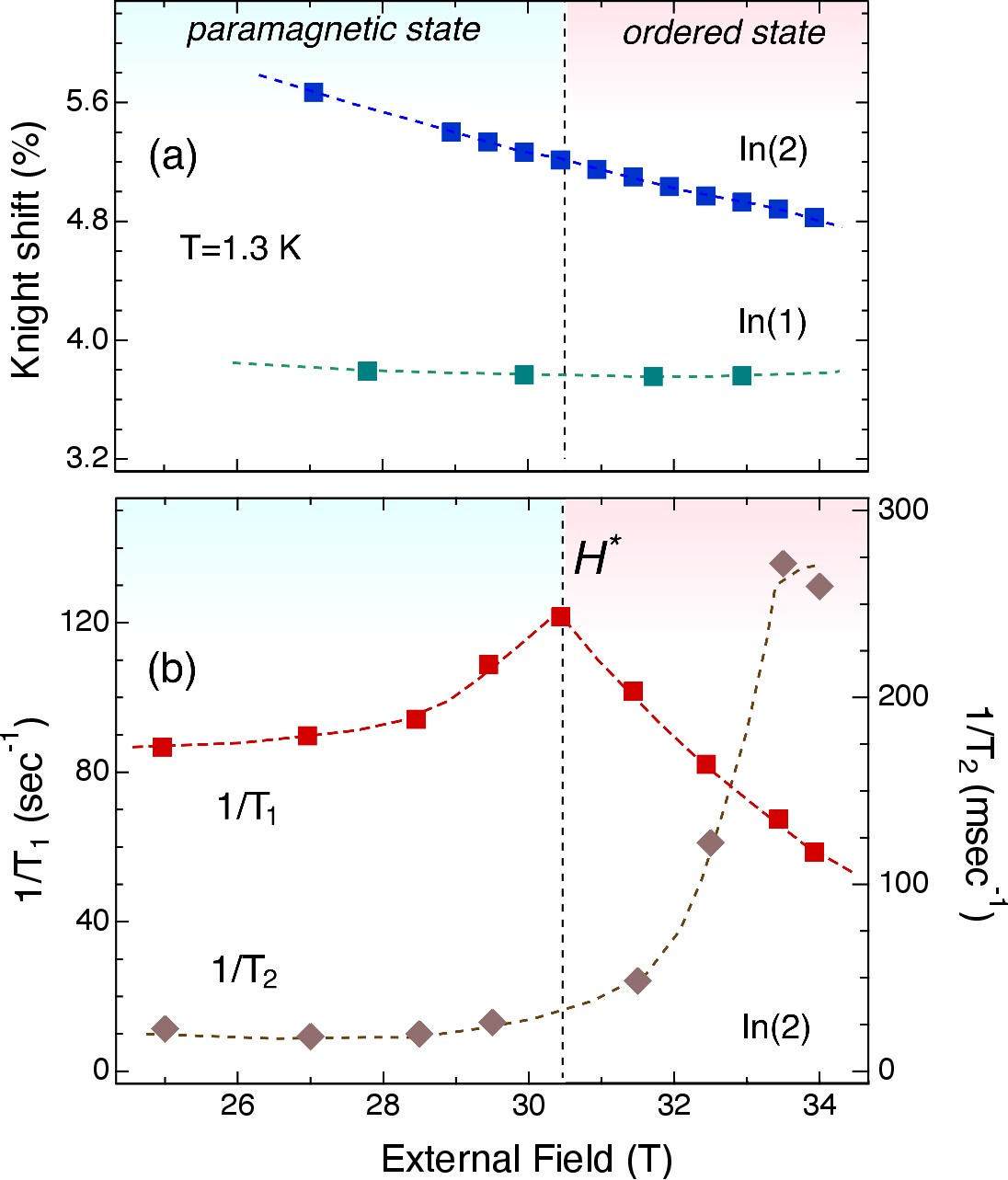}
\end{center}
\vspace{-2mm}
\caption{(a) Field dependence of the Knight shift for In(1) and In(2) sites and (b) inverse relaxation times for the In(2) site. The dashed line indicates the transition field $H^*$.}
\vspace{-4mm}
\label{f4}
\end{figure}

In Fig.~\ref{f4}(a), we present the magnetic field dependence of the NMR Knight shift, $K(H)$, for both In(1) and In(2) sites, measured in the field range from 27 to 34~T. For both sites, no anomaly is observed at $H^*$, consistent with the magnetization data $M(H)$. Interestingly, $K(H)$ exhibits a continuous decrease for In(2), while remaining nearly constant for In(1). This behavior can be related to a decrease of the out-of-plane transferred hyperfine coupling between Ce and In(2) with increasing $H$. In Ce-115 systems, the In(2) electrons are known to be more strongly hybridized with the Ce 4$f$ electrons~\cite{Shim2007,Haule2010}. Furthermore, changes in the shape of the 4$f$ wave function govern the strength of this hybridization and, therefore, the degree of itinerancy or localization of the 4$f$ electrons~\cite{Willers2015,Menegasso2021}. The observed reduction of the out-of-plane hybridization suggests that the 4$f$ wave function becomes increasingly flattened under magnetic fields, likely due to the field-induced mixing of the first excited state $\Gamma_7^1$ and the ground state $\Gamma_7^2$~\cite{Lesseux2020}. Consequently, a more localized character of the 4$f$ electrons is expected at high magnetic fields.

In contrast to the $K(H)$, distinct anomalies are clearly identified in the field dependence of $1/T_1$ and $1/T_2$ at $H^*$, as shown in Fig.~\ref{f4}(b). Specifically, $1/T_1$ shows a clear peak at $H^*$, and then continuously decreases above $H^*$. This behavior can be understood as a consequence of the critical slowing down of spin-fluctuations, followed by the opening of an excitation gap above $H^*$. Therefore, this is a strong signature of a magnetic phase transition at $H^*$. A strong increase in $1/T_2$ above $H^*$ implies the emergence of slowly fluctuating internal fields at the In sites within the ordered phase. As 1/$T_1$ is decreasing at the same time, these fluctuations are necessarily longitudinal.

Now, we discuss a possible origin of the field-induced order. Our high-field NMR results reveal that (i) the phase transition affects the spin dynamics, (ii) no large static internal fields appear along the $c$ axis at either In(1) or In(2) sites, and (iii) the EFG parameters with the local tetragonal symmetry at the In(1) site (i.e., $\eta=0$) are preserved in the ordered state. The result (i) indicates that the phase transition is accompanied by the ordering of Ce 4$f$ spin moments. On the other hand, (ii) narrows down potential propagation vectors in the ordered state, and (iii) denies the possibility of any structural transition or anisotropic charge ordering.

In magnetically ordered states, internal fields at nonmagnetic ligand sites arise from the spin density distribution of magnetic ions through dipolar and transferred hyperfine (HF) interactions. The transferred HF interaction emerges from the orbital hybridization effect. Since the induced magnetic field at a ligand site does not break the symmetry of the magnetic sublattice, possible directions for the internal field at a nonmagnetic ligand site can be deduced based on symmetry analysis~\cite{Demuynck2000,Ohama2005,Kambe2007}. Thus, it is possible to construct an invariant form for HF interactions at each ligand site \cite{Sakai1997,Sakai2005,Sakai2007,Kiss2008}. For In(2) sites in the 115 structure, the invariant form of HF interactions was derived by Kiss and Kuramoto \cite{Kiss2008}.

Based on this form, in the following, we derive the constraints on the magnetic ordering. We recall that only the local fields that are parallel to the applied magnetic field (here $H \parallel c$) are observable in an NMR spectrum. For our structure, these fields are [39]
\begin{flalign}
\;\;\;\;\;\;\;&h_c(\textrm{In}(2_{a,b})) = c_{\perp} \sin(aQ_{a,b}/2)\mu_{a,b}(\textbf{Q}) \nonumber &\\
&\;\;\;\;\;\;\;\;\;\;\;\;\;\;\;\;\;\;\;\;\;\;+ c_c \cos(aQ_{a,b}/2) \mu_c(\textbf{Q})& \\
&h_c(\textrm{In}(1)) = c_1 \cos(aQ_a/2) \cos(aQ_b/2) \mu_c(\textbf{Q}),&
\end{flalign}
where $\textrm{In}(2_{a,b})$ refers to the $\textrm{In}(2)$ sites that are in the $ac$ and $bc$ planes, $\mu_{a,b,c}(\textbf{Q})$ denote the components of the ordered Ce moments, and $c_{\perp}$, $c_c$, and $c_1$ are unknown constants~\footnote{Here $a$ and $b$ are treated as inequivalent to allow for magnetic states with in-plane moments that break tetragonal symmetry, although they are equivalent ($a=b$) in the tetragonal paramagnetic phase.}. From (ii), we know that no local fields are observed, $h_c(\textrm{In}(2_{a,b}))$ = 0 = $h_c(\textrm{In}(1))$. Any nonzero ordered moment component must thus be canceled by the zero value of the above given ``form factors.'' This provides the following constraints on the $\textbf{Q}$ values:
\begin{flalign}
\;\;\;\;\;\;\;\;&\mu_a \neq 0\textrm{~~}\Rightarrow\textrm{~~}Q_a = 0 & \\
&\mu_b \neq 0\textrm{~~}\Rightarrow\textrm{~~}Q_b = 0 & \\
\label{mu_c}
&\mu_c \neq 0\textrm{~~}\Rightarrow\textrm{~~}Q_a = Q_b = 1/2\times2\pi/a. &
\end{flalign}
That is, if the moments are oriented within the plane, their in-plane order must be either striped (columnar) when aligned along one of the crystal axes, or ferromagnetic otherwise. However, such in-plane stripe or ferromagnetic orders would break the tetragonal site symmetry at the In(1) sites, and the corresponding $\eta$ value would therefore become nonzero due to the magnetoelastic effect. This is inconsistent with (iii). Furthermore, as will be shown later, these in-plane stripe or ferromagnetic orders are incompatible with the dHvA results, which revealed that only the largest $\beta_1$ frequency disappears above the transition. An incommensurate spiral magnetic order with in-plane moments observed in CeRhIn$_5$ \cite{Curro2000,Bao2000,Fobes2017} is also ruled out by (3) and (4).

If the moments are oriented perpendicular to the plane, their in-plane order must be fully AFM. The corresponding propagation vector is therefore either $\mathbf{Q} = (1/2, 1/2, 0)$ or $\mathbf{Q} = (1/2, 1/2, 1/2)$. Interestingly, the propagation vector $\mathbf{Q} = (1/2, 1/2, 1/2)$ with $\mu_\mathrm{ord} \parallel [0, 0, 1]$ was observed in 10\% Cd-doped CeIrIn$_5$ by elastic neutron scattering~\cite{Beauvois2020}. In general, AFM structures with moments aligned along the field direction are energetically unfavorable. However, in Ce-based heavy-fermion systems, the direction of the ordered moment is often governed by the crystal electric field anisotropy of the Ce 4$f$ ground state rather than by the Zeeman energy alone (see, e.g.\,Ref.\,\cite{Moll2017}). In CeIrIn$_5$, the $c$ axis remains the easy magnetization axis even at very high fields \cite{Takeuchi2001}. Furthermore, although the static local fields cancel at the nuclear sites in these AFM structures, fluctuations of the ordered moments—either in magnitude or direction—can lead to incomplete cancellation of the local field along the $c$ axis, thereby producing slow longitudinal fluctuations, as detected in $1/T_2$. In the following, we further narrow down the possible propagation vectors by considering the results of previous dHvA measurements~\cite{Aoki2016}.

\begin{figure}[htb]
\begin{center}
\includegraphics[width=8.4cm,keepaspectratio]{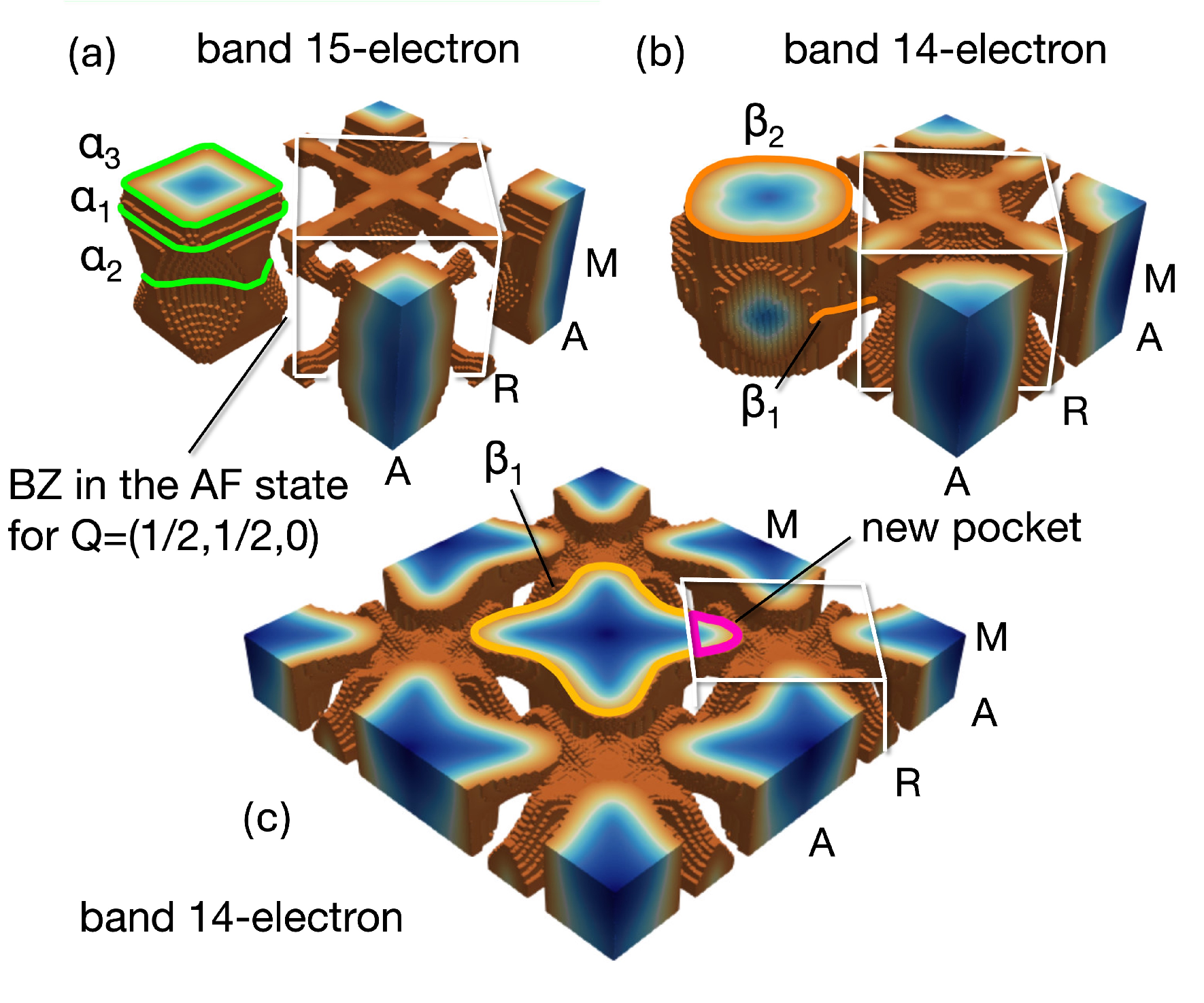}
\end{center}
\vspace{-4mm}
\caption{Calculated major FSs in the paramagnetic state of CeIrIn$_5$~\cite{Suppl}. They consist of (a) the $\alpha$ sheet originating from the electron band 15 and (b) the $\beta$ sheet originating from the electron band 14. Here, the band numbers follow the notation used in previous papers~\cite{Haga2001}. (c) Cross section of the $\beta$ sheet illustrating the largest $\beta_1$ orbit. Magnetic Brillouin zone for the AFM state with $\mathbf{Q} = (1/2, 1/2, 0)$ is also shown.}
\label{f5}
\vspace{-4mm}
\end{figure}

Any AFM order results in magnetic Brillouin zones (BZs) that are smaller than the crystallographic BZ. Consequently, the FSs are also modified due to the band folding. For a magnetic structure with $\mathbf{Q} = (1/2, 1/2, 0)$ [Fig.\,1(b)], the topology of the magnetic BZ is the same as that of its nonmagnetic counterpart, but its cross-sectional area is reduced by a factor of 2, as illustrated in Fig.~\ref{f5}. This modification of the BZ does not impact the $\alpha$ sheet of the FS, which fits into the smaller magnetic BZ. The $\beta_2$ orbit of the $\beta$ sheet centered at the $A$ point in the momentum space is not impacted either, as can be seen in Fig.~\ref{f5}(b). However, the $\beta_1$ orbit centered at the $M$ point does not fit into the magnetic BZ. The corresponding dHvA frequency should, therefore, disappear in the magnetically ordered state. On the other hand, a new small FS pocket should emerge due to the band folding [see Fig.~\ref{f5}(c)]. Such a pocket results in a new low dHvA frequency. This is exactly what was observed in previous high-field dHvA measurements~\cite{Aoki2016}.

In contrast, for $\mathbf{Q} = (1/2, 1/2, 1/2)$, the topology of the magnetic BZ differs from that of the paramagnetic state. Consequently, all FS sheets would be reconstructed, leading to a complete change in the dHvA frequencies. Regarding possible in-plane orders, the only allowed propagation vectors from Eqs.\,(3) and (4) are  $\mathbf{Q} = (0, 1/2, 0)$, $(1/2, 0, 0)$, $(0, 1/2, 1/2)$, $(1/2, 0, 1/2)$ and $(0, 0, 1/2)$. The corresponding magnetic BZs either give rise to an additional extremal cross section of the $\alpha$ sheet or affect the $\beta_2$ orbit. This results in either emergence of a new dHvA frequency of the order of the $\alpha$ frequencies or disappearance of the $\beta_2$ frequency above the transition. Neither was observed in the experiment~\cite{Aoki2016}. Therefore, only the AFM order with the propagation vector $\mathbf{Q} = (1/2, 1/2, 0)$ can account for the experimental observations~\cite{Aoki2016}. Within this scenario, the $\beta_1$ orbit could still be observed via magnetic breakdown if the applied field is sufficiently strong. It would be intriguing to test this hypothesis experimentally by performing dHvA measurements at even higher magnetic fields.


In summary, we performed low-temperature high-field NMR measurements on a single crystal of CeIrIn$_5$. We observed a clear change in the spin dynamics, both in $1/T_1$ and $1/T_2$, at $H^* \simeq$~30.5~T, demonstrating a transition of a magnetic origin. Furthermore, as the NMR spectrum does not change across $H^*$, the transition is most likely to be AFM with a limited set of magnetic structures. Taking into account previous high-field dHvA results together with band-structure calculations further reduces the choice to an AFM spin alignment along the $c$-axis with the propagation vector $\mathbf{Q} = (1/2, 1/2, 0)$. If our hypothesis is correct, this makes CeIrIn$_5$ outstanding in two aspects. First, contrary to other heavy-fermion compounds, here magnetic field induces rather than suppresses the AFM transition. Second, above the transition, the magnetic moments are antiferromagnetically aligned along the field direction. High-field neutron diffraction measurements are required to definitely confirm our hypothesis.

\begin{acknowledgments}
This work was supported by LNCMI-CNRS, a member of the European Magnetic Field Laboratory (EMFL), ICC-IMR, the REIMEI Research Program of JAEA, and JSPS KAKENHI Grants No. JP20KK0061, No. JP22H04933, No. JP23K25827, No. JP24H01641, No. JP24K00581, JP24K00587, No. JP24K00588, No. JP24KK0062, No. JP25K00947, and No. JP25K21684.
\end{acknowledgments}

\bibliography{CeIrIn5_NMR}

\clearpage

\onecolumngrid

\renewcommand{\figurename}{Figure S\hspace{-0.1cm}}

\setcounter{figure}{0}

\begin{center}
\textbf{Supplementary Information for ``Field-induced antiferromagnetic transition in CeIrIn$_5$"}
\end{center}

\section{Low-field NMR spectrum and line simulations}

The upper panel of Fig.~S1 shows the frequency-swept NMR spectrum measured on a single crystal of CeIrIn$_5$ at a field of 12.1~T ($H \parallel c$) and a temperature of 1.73~K. The same single crystal was used for the high-field NMR experiments described in the main text. The spectrum consists of 18 (9$\times$2) peaks originating from the In1 and In2 sites. The peaks are very sharp, attesting to the high quality of our single crystals.

\begin{figure}[h]
\includegraphics[width=12cm,keepaspectratio]{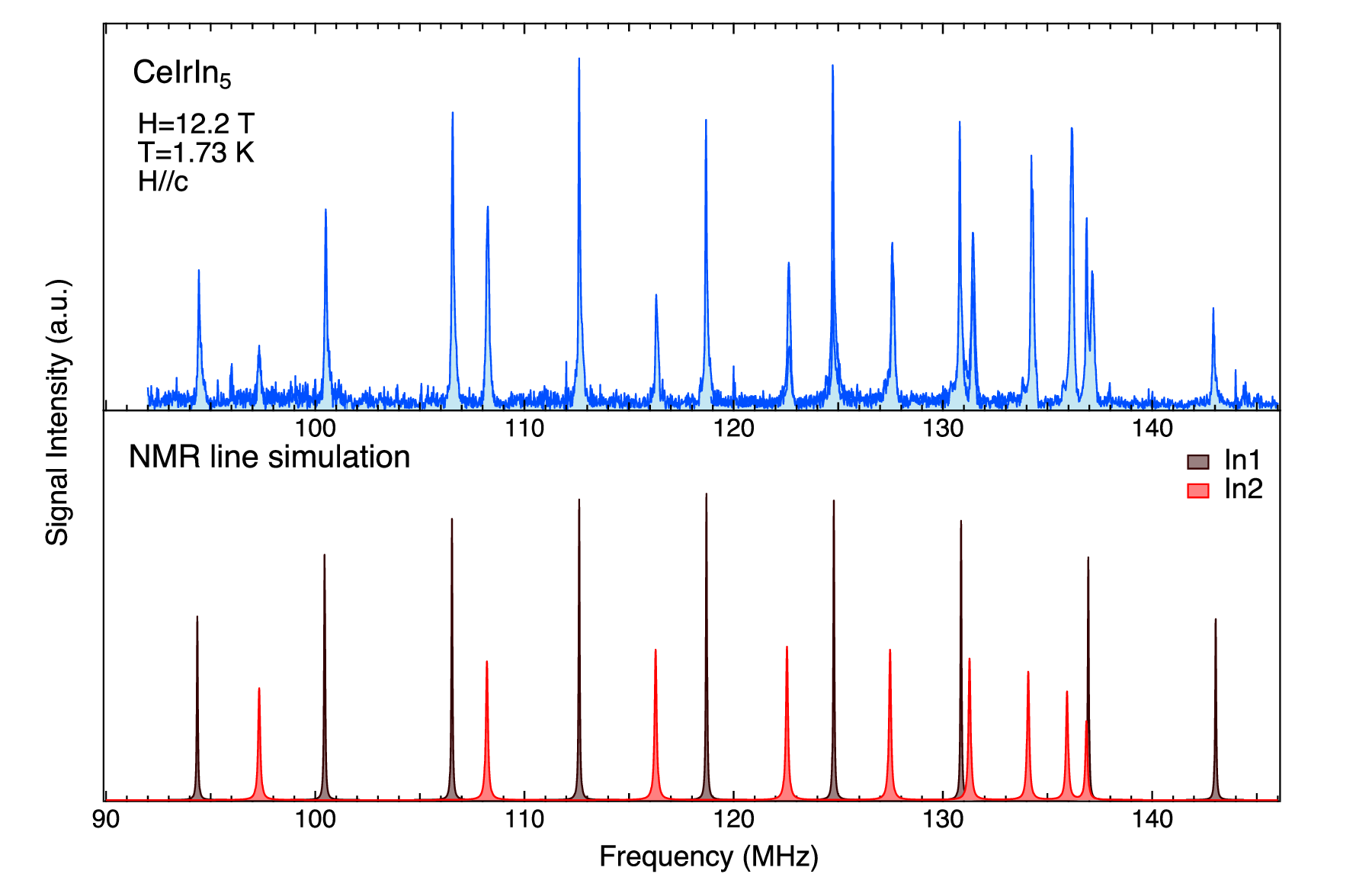}%
\caption{Upper panel: frequency-swept NMR spectrum of a single crystal of CeIrIn$_5$ at 12.1 T ($H \parallel c$) and 1.73 K. Lower panel: result of the lines simulation (see text for details). The principal axes of their EFG tensors are illustrated in Fig. 1 of the main text.}
\end{figure}

To evaluate the EFG parameters from the spectrum, we have carried out numerical simulations of the spectra using a diagonalized total Hamiltonian matrix, which consists of the Zeeman term and the quadrupolar term,
\begin{equation}
\mathcal{H} \rm= \mathcal{H}_{\rm z} + \mathcal{H}_{\rm q}= -\gamma_{\rm N}\hbar(1+\bvec{K})\bvec{I} \cdot \bvec{H}
+\frac{\hbar \nu_{\rm Q}}{6}\Bigl\{ (3I_{\rm Z}^2-\bvec{I}^2)+\frac{1}{2}\eta(I_{\rm +}^2+I_{\rm -}^2)\Bigr\}.
\label{e1}
\end{equation}
Here, $K$ is the Knight shift, $\nu_Q$ is the quadrupolar frequency, and $\eta$ is the asymmetry parameter of the EFG.
The best fit to the experimental data was obtained with parameters, $K=4.28$\%, $\nu_{\rm Q}=6.083$ MHz, and $\eta=0$ for the In1 sites, and $K=7.54$\%, $\nu_{\rm Q}=18.167$ MHz, and $\eta=0.465$ for the In2 sites, respectively. The estimated values of $\nu_{\rm Q}$ and $\eta$ agree with the values obtained from previous NMR and NQR measurements \cite{Zheng2001,Kohori2001}.

\section{Band-structure calculation}

The electronic band structure was calculated within the framework of the full-potential linearized augmented plane wave (FLAPW) method using the local density approximation (LDA) for the exchange-correlation potential. The spin–orbit interaction is included selfconsistently for all valence electrons in a second variational procedure. In the calculations, we included Ce 5d/6s/4f, Ir 5d/6s, and In 4d/5s/5p orbitals as valence states and Ce 5$s$/5$p$, and Ir 5p/4f orbitals as semi-core states. The calculation was performed for the nonmagnetic ground state by assuming the presence of the time-reversal symmetry.


\end{document}